# Acquisition-Independent Deep Learning for Quantitative MRI Parameter Estimation using Neural Controlled Differential Equations

Daan Kuppens[1,2*], Sebastiano Barbieri[3,4], Daisy van den Berg[1,5], Pepijn Schouten[1], Harriet C. Thoeny[6,7], Myrte Wennen[1], Oliver J. Gurney-Champion[1,2]

[1] Department of Radiology and Nuclear Medicine, Amsterdam University Medical Center, Amsterdam, The Netherlands

[2] Imaging and Biomarkers, Cancer Center Amsterdam, Amsterdam, The Netherlands

[3] Queensland Digital Health Centre, University of Queensland, Brisbane, Australia

[4] Centre for Big Data Research in Health, UNSW Sydney, Sydney, Australia

[5] Department of Biomedical Engineering and Physics, Amsterdam University Medical Center location University of Amsterdam, Amsterdam, The Netherlands

[6] University Teaching and Research Hospital, University of Fribourg, Fribourg, Switzerland

[7] Department of Urology, Inselspital, University of Bern, Switzerland

* Corresponding author: d.kuppens@amsterdamumc.nl

## Abstract

Deep learning has proven to be a suitable alternative to least-squares (LSQ) fitting for parameter estimation in various quantitative MRI (QMRI) models. However, current deep learning implementations are not robust to changes in MR acquisition protocols. In practice, QMRI acquisition protocols differ substantially between different studies and clinical settings. The lack of generalizability and adoptability of current deep learning approaches for QMRI parameter estimation impedes the implementation of these algorithms in clinical trials and clinical practice. Neural Controlled Differential Equations (NCDEs) allow for the sampling of incomplete and irregularly sampled data with variable length, making them ideal for use in QMRI parameter estimation. In this study, we show that NCDEs can function as a generic tool for the accurate prediction of QMRI parameters, regardless of QMRI sequence length, configuration of independent variables and QMRI forward model (variable flip angle T1-mapping, intravoxel incoherent motion MRI, dynamic contrast-enhanced MRI). NCDEs achieved lower mean squared error than LSQ fitting in low-SNR simulations and in vivo in challenging anatomical regions like the abdomen and leg, but this improvement was no longer evident at high SNR. NCDEs reduce estimation error interquartile range without increasing bias, particularly under conditions of high uncertainty. These findings suggest that NCDEs offer a robust approach for reliable QMRI parameter estimation, especially in scenarios with high uncertainty or low image quality. We believe that with NCDEs, we have solved one of the main challenges for using deep learning for QMRI parameter estimation in a broader clinical and research setting.

1. **Introduction**

Imaging biomarkers, such as quantitative MRI (QMRI) parameters, offer accessible, cost-effective, reproducible and non-invasive tools to assist in lesion detection, characterization and treatment monitoring of various pathologies improving patient care. QMRI techniques produce parameters used to assess tissue morphology, biology, and function. Proven useful clinical applications include oncological monitoring (Chauvie et al., 2023; O'Connor et al., 2017) and the imaging of stroke (Albers, 1998), detection of myocardial abnormalities (Manfrini et al., 2021; Taylor et al., 2016) and iron overload (Liden et al., 2021). Commonly studied QMRI techniques include the intra-voxel incoherent motion (IVIM MRI) model for diffusion-weighted imaging (DWI), T1-relaxometry using variable flip angle (VFA T1-mapping) and dynamic contrast-enhanced MRI (DCE-MRI).

In QMRI, tissue properties are estimated from a series of MRI data using biophysical models that relate the measured MRI signal to the underlying tissue properties via QMRI parameters. Conventionally, such parameters are estimated with least squares fitting (LSQ) to retrieve the QMRI parameters from MR images with different contrast weightings. LSQ fitting is an iterative process that minimizes the sum of squared differences between observed MRI data and the reconstructed signal curve. Although LSQ fitting is a reliable estimator of physiological parameters when SNR is high, it does have significant limitations. When SNR is low, the combined effect of ignoring spatial information, noisy signal curves and non-convex objective functions leads to high variance in the QMRI parameter estimates (Barbieri et al., 2016; Neil, 1993; While, 2017) The compromised repeatability and accuracy of parameter estimates form a crucial hurdle for the clinical application of QMRI techniques (Kurland et al., 2012; Rosenkrantz et al., 2015)

Recent work demonstrated that deep learning, with its capacity to learn nonlinear mappings, is a suitable alternative to LSQ fitting for estimating parameters in many QMRI models. In particular, it improves accuracy and precision, yields faster parameter estimation and reduces day-to-day variation in patients (Barbieri et al., 2020; Bliesener et al., 2020; Gurney-Champion et al., 2022; Kaandorp et al., 2021; Ottens et al., 2022; Ulas et al., 2018). However, in contrast to LSQ fitting these deep learning implementations are not robust to different MR acquisition protocols, as they are dependent on the input being either a fixed set of input signals (for e.g. fully connected networks (Bliesener et al., 2020; Kaandorp et al., 2021; Ulas et al., 2018) and convolutional networks (Huang, 2022; Vasylechko et al., 2022)) or a series of regularly sampled signals (for recurrent neural networks (Ottens et al., 2022)). In practice, QMRI acquisition protocols differ substantially between studies and clinical settings (Ljimani et al., 2020), and acquisition settings do not typically follow equidistant sampling patterns (e.g. flip angles 2, 5, 10, 25°). The lack of generalizability and adoptability of current deep learning approaches for QMRI parameter estimation impedes the implementation of these algorithms in clinical trials and clinical practice. Hence, an acquisition-independent approach is crucial for implementation of deep learning for QMRI parameter estimation in the clinical workflow.

In parallel, a group of Neural Ordinary Differential Equations has been developed as machine learning methods that approximate system dynamics in continuous time by training a neural network to learn the underlying differential equation (Chen, 2018). Neural Controlled Differential Equations (NCDEs) build on this framework by incorporating incoming data to control the learnt differential equation with observations and thereby creating an explicit dependence of the output on the learnt system dynamics and the input series (Kidger, 2020). NCDEs allow for the sampling of incomplete and irregularly sampled data with variable length, making them ideal for use in QMRI parameter estimation.

We hypothesize that NCDEs can function as generalizable acquisition-independent networks for QMRI parameter estimation. This would solve the abovementioned shortcomings and pave the way for the integration of deep learning for QMRI in the clinical workflow.

Our main contributions are:

- We overcome the limitation that neural networks are specific to MR acquisition protocols by implementing NCDEs for QMRI parameter estimation.
- We demonstrate NCDE performance on simulated data for VFA T1-mapping, IVIM MRI and extended Tofts-Kety DCE-MRI.
- We demonstrate NCDE performance on in vivo VFA T1-mapping.
- The experimental results demonstrate the superiority of NCDEs compared to conventional LSQ fitting for QMRI parameter estimation, both in simulation and in vivo.

## 2. **Methods**

All analyses were performed using Python (v3.8, Python Software Foundation) and PyTorch (v1.13.1, PyTorch Foundation).

### 2.1 Quantitative MRI

QMRI models allow us to assess tissue properties by using biophysical models to describe the MRI signal intensity as a function of a changing independent variable, such as flip angle (FA) in VFA T1-mapping, diffusion weighting (b-value) in IVIM MRI and time (t) in DCE-MRI.

**In VFA T1-mapping**, the longitudinal spin relaxation time ($T1$) is estimated by acquiring multiple spoiled gradient-echo readouts, each with different excitation flip angles ($FA$). Consequently, the signal ($s$) at $FA$ depends on $T1$, the repetition time ($TR$), and the magnetization at thermal equilibrium ($s_0$) (Christensen, 1974; Gupta, 1977):

$$s(FA, TR, s_0) = s_0 \frac{1 - \exp\left(-\frac{TR}{T1}\right)}{1 - \cos(FA) \exp\left(-\frac{TR}{T1}\right)} \sin(FA). \tag{1}$$

**In IVIM MRI**, diffusion-weighted gradients are used to sensitize the MRI signal to incoherent motion. By varying the strength and duration of the diffusion-weighted gradients, we are able to separate the effect of incoherent motion from diffusion with a bi-exponential model where the signal ($s$) intensity at diffusion weighting ($b$) depends on the diffusion coefficient ($D$), which indicates tissue cell density and structural integrity, the pseudo-diffusion coefficient ($D^*$) and perfusion fraction ($f$), which reflect tissue vascularization and perfusion, and the baseline signal intensity ($s_0$) :

$$s(b, s_0) = s_0((1 - f) \exp(-bD) + f \exp(-bD^*)). \tag{2}$$

**In DCE-MRI**, tissue $T1$ is measured over time after bolus administration of a contrast agent, gadolinium. Gadolinium shortens the tissue $T1$, allowing the tissue gadolinium concentration over time to be estimated. In turn, the pharmacokinetics of gadolinium reflects the tissue perfusion and blood vessel permeability (Khalifa et al., 2014; Sourbron and Buckley, 2013). In the extended Tofts-Kety Model for DCE-MRI, tissue gadolinium concentration ($C_t$) at time $t$ depends on the reflux rate ($k_e$), fractional volume of extravascular extracellular space ($v_e$), fractional volume of plasma ($v_p$) and the concentration of gadolinium in blood plasma ($C_p$) (Kety, 1951; Tofts, 1999):

$$C_t(t) = v_p C_p(t) + k_e v_e \int_0^t C_p(\tau) \exp(-k_e(t-\tau)) d\tau.$$

(3)

For consistency between QMRI forward models throughout the text, we will refer to $C_t$ as $s$. For the extended Tofts-Kety DCE-MRI model, we utilized the implementation and fitting routines from the Python package OG_MO_AUMC_ICR_RMH_NL_UK (Orton et al., 2008), available on the OSIPI GitHub repository (van Houdt et al., 2024). In all described experiments, $C_p$ remained fixed (for detailed information, see Table S4).

## 2.2 NCDE model and training

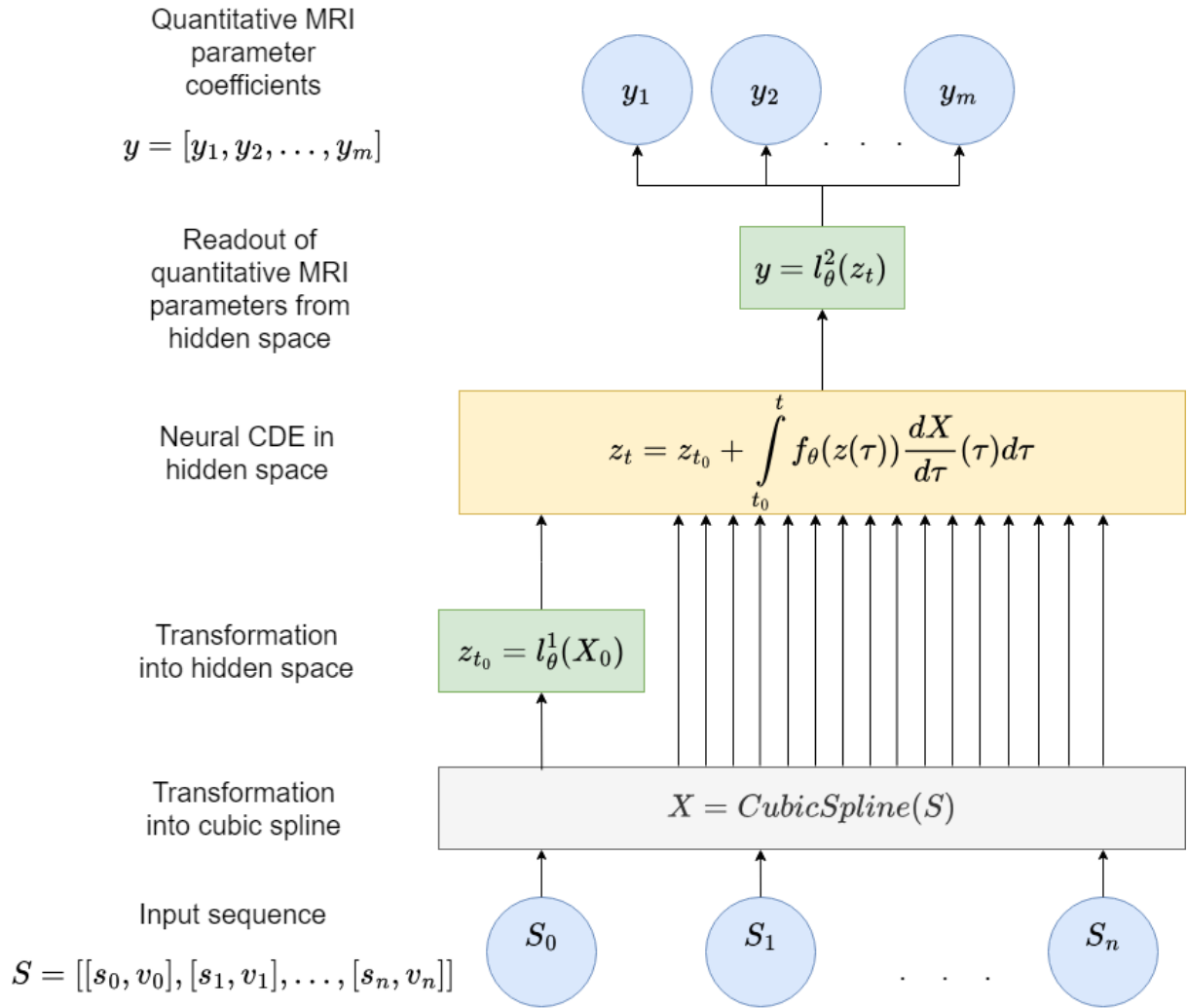

*Fig. 1: schematic representation of an NCDE for QMRI parameter estimation. The NCDE is composed of three neural networks: $l_\theta^1$, $f_\theta$, and $l_\theta^2$. The input sequence $S$ can have arbitrary length and irregular sampling intervals. The length of output vector $y$*

*corresponds to the number of predicted QMRI parameter coefficients. X is a twice continuously differentiable cubic spline, whose knots are the elements of S. z denotes the hidden state.*

The proposed approach assumes the existence of a hidden space where changes in the measured signal curve can be described by a controlled differential equation. NCDEs (Fig. 1) learn a parameterized mapping from the input sequence $S$ to output $y$, using three fully connected neural networks ($l_\emptyset^1$, $f_\emptyset$ and $l_\emptyset^2$). The input sequence ($S$) contained the signal intensities ($s$, either a series of MRI signals for VFA T1-mapping and IVIM MRI or a derived gadolinium concentration curve for DCE-MRI) along with corresponding values of the independent variable ($v$, either the flip angle for VFA T1-mapping, b-value for IVIM MRI or time for DCE-MRI). The input sequence defines the knots of a twice continuously differentiable cubic spline ($X$), representing an approximation of the continuous underlying process which is observed through $S$. $l_\emptyset^1$ maps the first element of the input sequence ($X_0$) to the initial value in hidden space ($z_{t_0}$). The hidden state ($z$) evolves within the hidden space according to the product of $f_\emptyset(z)$ and the derivative of the input sequence with respect to an auxiliary variable ($\frac{dX}{d\tau}$). $f_\emptyset$ represents a learned vector field in hidden space, the auxiliary variable functions as a placeholder for the independent variable. The hidden state at $t$ ($z_t$) is then obtained by solving the controlled $f_\emptyset$ over the distance between $t_0$ and $t$, following:

$$z_t = z_{t_0} + \int_{t_0}^t f_\emptyset\big(z(\tau)\big)\frac{dX}{d\tau}(\tau)d\tau. \tag{4}$$

$l_\emptyset^2$ reads out the final hidden state and maps it to an output vector ($y$) containing coefficients of the estimated QMRI parameters.

In our implementation, we kept the architecture similar to the original implementation of NCDEs (Kidger, 2020). As $l_\emptyset^1$ merely initializes the hidden space, it is a single-layer perceptron. $f_\emptyset$ is a multi-layer perceptron consisting of 6 layers with variable widths. $l_\emptyset^2$ contains multiple parallel multi-layer perceptrons, one for each QMRI parameter (Kaandorp et al., 2021). For further details on the architecture of the NCDEs, see Table S1.

We trained an NCDE to map the input sequence to M output parameters, where M=2 for VFA T1 mapping, M=4 for IVIM MRI and M=4 for DCE-MRI. Results on parameters required for modelling but not representing relevant physiological quantities ($s_0$ in VFA T1-mapping and IVIM MRI, $d\tau$ in DCE-MRI) are not presented. To constrain the estimates, output parameters were rescaled to pre-specified ranges [$y_{i,min}$, $y_{i,max}$] using:

$$\hat{y}_i = y_{i,min} + \sigma(\hat{y}_{i,coeff})(y_{i,max} - y_{i,min}), \tag{5}$$

where $\hat{y}_i$ is the estimate for $y_i$ ($i$ has range: [1, n]), $\hat{y}_{i,coeff}$ is the NCDE output and $\sigma$ is the sigmoid function. The ranges for $y_{i,min}$ and $y_{i,max}$ are described per QMRI parameter in table 1. Based on the predicted QMRI parameters together with the set of measured values for the independent variable (flip angle in VFA T1-mapping, b-value in IVIM MRI and time in DCE-MRI), a signal curve $S_{pred}$ is predicted according to Eqs. (1-3).

For each QMRI forward model, denoted in Eqs. (1-3), a dedicated NCDE is trained by minimizing a loss ($L$), which is a linear combination of a physics-informed loss (Barbieri et al., 2020) and a supervised loss, following:

$$L = \sum_{i=1}^n ((\hat{y}_{i,coeff} - y_{i,coeff})^2 + \frac{1}{len(s_{i,input})}\sum(s_{i,input} - s_{i,pred})^2). \tag{6}$$

When trained on in vivo data, where no supervisory coefficients for the estimated QMRI parameters are available, the combined loss in Eq. (6) reduces to the physics-informed loss.

## 2.3 Data

### 2.3.1 Simulations

To show the generalizability of NCDEs in the context of QMRI, we trained three different NCDE models to estimate IVIM MRI, VFA T1-mapping and DCE-MRI parameters.

Per QMRI model, 1,000,000 training signal (for IVIM MRI and VFA T1-mapping) or concentration (for DCE-MRI) curves were simulated using the forward model as described in Eqs. (1-3). Individual QMRI parameters were sampled from a uniform distribution with ranges as described in Table 1. Independent variables (flip angle, b-value or time) were sampled for each training signal as described in Table 2. The sampling scheme reflects clinical imaging protocols, while maintaining a source of variability in both sequence length and values of the independent variable. This allowed training and validating NCDEs on a large range of representative configurations of independent variables.

For VFA T1-mapping and IVIM MRI, noise was added to the simulated signal curves to make the data follow a Rician distribution similar to MRI magnitude images. The signal-to-noise ratio was set to a predefined value ($SNR_{predefined}$) and the noisy signal ($s_{noisy}$) followed:

$$s_{noisy} = \sqrt{(s + N(0, \frac{s_0}{SNR_{predefined}}))^2 + N(0, \frac{s_0}{SNR_{predefined}})^2}, \qquad (7)$$

where $s_0$ refers to the signal at $b = 0$ $^s/_{mm^2}$ for IVIM MRI and the magnetization at thermal equilibrium for VFA T1-mapping. Here, $N(\mu, SD)$ represents random sampling from a normal distribution with mean μ and standard deviation SD. For DCE-MRI, the measurements reflect gadolinium concentration instead of an MR signal and therefore Gaussian noise instead of Rician noise was added to the generated curves following:

$$s_{noisy} = s + N(0, \frac{s_{max}}{SNR_{predefined}}). \qquad (8)$$

In DCE-MRI, $s_{max}$ refers to the max signal of the signal curve.

| Quantitative MRI model | Parameter | Range |
|---|---|---|
| VFA T1-mapping | $T1$ | $100 - 3000$ ms |
| IVIM MRI | $D$ | $0.00035 - 0.003$ $mm^2/_s$ |
| | $D^*$ | $0.05 - 0.1$ $mm^2/_s$ |
| | $f$ | $0.03 - 0.25$ |
| DCE-MRI | $k_e$ | $0.1 - 2$ min$^{-1}$ |
| | $v_e$ | $0.01 - 0.7$ |
| | $v_p$ | $0.001 - 0.05$ |

*Table 1: description of QMRI parameter ranges used in simulated data during training and evaluation. QMRI parameters were randomly sampled from a uniform distribution with these ranges.*

| Quantitative MRI model | Sequence length | Independent variable | Range of independent variable | Sampling protocol | SNR range |
|---|---|---|---|---|---|
| VFA T1-mapping | [3, 8] | Flip angle | [0, 30]° | For $N$ flip angles: <br> 1. Sample flip angle of 1° <br> 2. Randomly sample 1 flip angle in range [3,10]° <br> 3. Randomly sample 1 flip angle in range [15,30]° <br> 4. Equidistantly sample N-3 flip angles between 1° and maximum flip angle | [50, 300] |
| IVIM MRI | [4, 10] | b-value | [0, 800] $s/mm^2$ | For $N$ b-values: <br> 1. Sample b-value of 0 $s/mm^2$ <br> 2. Randomly sample $\lceil \frac{\lceil(\frac{N-1}{2})\rceil}{2} \rceil$ b-values <50 $s/mm^2$ <br> 3. Randomly sample $\lfloor \frac{\lceil(\frac{N-1}{2})\rceil}{2} \rfloor$ b-values in range [50,99] $s/mm^2$ <br> 4. Randomly sample $\lfloor(\frac{N-1}{2})\rfloor$ b-values [100, 800] $s/mm^2$ | [5, 40] |
| DCE-MRI | [15, 20] | Time | [0, 240] $s$ | For $N$ timepoints: <br> 1. Randomly sample temporal resolution in range [8,12] s | [5, 40] |

*Table 2: description of the protocol for sampling the independent variables and simulating the corresponding signal curves. These simulation procedures have been used during training and evaluation. Even though the denoted range in SNR appears vastly different among the three QMRI simulations, this is caused by differences in the SNR definition. The ceiling operator ($\lceil x \rceil$) rounds x up to the nearest integer, the floor operator ($\lfloor x \rfloor$) rounds x down to the nearest integer.*

### 2.3.2    In vivo VFA T1-mapping

VFA T1-mapping MRI data of the abdomen, leg and brain were acquired in 8 healthy volunteers using a 3T MRI scanner (Ingenia, Philips, Best, The Netherlands) at the Amsterdam UMC. By scanning three anatomical regions, a broad range of T1 values was covered in the experiments. In one volunteer, no data from the abdomen was acquired. In another volunteer, no data from the leg was acquired. The study was approved by the local medical ethics review committee. All participants provided written informed consent, and research was performed in accordance with the Declaration of Helsinki guidelines. To obtain a complete VFA T1-mapping dataset, T1-weighted 3D spoiled gradient echo MRI was acquired at 30 flip angles (1-30°). Detailed MRI acquisition settings are provided in Table S2.

Because in vivo analysis in all QMRI models was not feasible, we chose to conduct it in the QMRI model most suitable for dense sampling of the independent variable, VFA T1-mapping.

### 2.3.3    Preprocessing

Neural networks perform optimally for normalized signals, while in VFA T1-mapping and IVIM MRI signals are in arbitrary units. Therefore, we normalized the data by dividing the signal curve by an approximate $s_0$. For VFA T1-mapping, at low flip angles Eq. (1) can be rewritten to:

$$s_0 \approx \frac{s(FA)}{\sin(FA)}. \tag{9}$$

In IVIM MRI, $s_0$ was approximated by the signal at b=0 $s/mm^2$. For DCE-MRI modelling, the input signal is the contrast concentration in M (molarity), and is thus inherently normalized.

## 2.4 Experiments

### 2.4.1 Training and evaluation on simulation data

Three NCDE models (one for each QMRI forward model) were trained for 750 epochs, with each epoch consisting of 100 data batches of size 64. Each training set (separate for VFA T1-mapping, IVIM MRI and DCE-MRI) contained 1,000,000 simulated signal curves. A validation set was not used, as the training was based on a fixed number of epochs. These curves were generated with varying sequence lengths and configurations of independent variables, following the procedure outlined in Section 2.3.1 and Table 2. Each training used the Adam optimizer (Kingma, 2015). Learning rate started at $1 \cdot 10^{-4}$, and halved at epochs 250, 350, 450, 550 and 650. During training, an adaptive step solver adjusts the step size so that the error in the solution remains approximately equal to a predefined tolerance (Kidger, 2020). Absolute and relative error tolerance of the adaptive ODE solver to solve for the hidden state in Eq. (4) were set to 0.00001 and 0.001 respectively and halved at epochs 300, 400, 500 and 600.

QMRI parameter estimates were evaluated across different sequence lengths and SNR levels and compared between LSQ fitting and NCDEs. To achieve this, 1,000,000 curves per QMRI forward model (VFA T1-mapping, IVIM MRI and DCE-MRI) were generated with varying sequence lengths and configurations of independent variables, following the procedure outlined in Section 2.3.1 and Table 2. Since for simulations the ground truth QMRI parameter values are known, the error (as in Eq. (10)) was calculated using:

$$\text{error} = \hat{y} - y. \tag{10}$$

Where $\hat{y}$ denotes the estimated QMRI parameter and $y$ the ground truth QMRI parameter. The squared error was calculated using:

$$\text{squared error} = (\hat{y} - y)^2. \tag{11}$$

In addition, the mean, median, 25th and 75th percentiles of the error and squared error over all curves were calculated. The mean of the squared error functions as a measure of accuracy. The mean of the error functions as a measure of bias. The range between the 25th and 75th percentile (interquartile range, or IQR) functions as a measure of precision.

### 2.4.2    Training and evaluation on in vivo VFA T1-mapping data

Four NCDE models were trained to estimate VFA T1-mapping parameters using 4-fold cross-validation, with each fold utilizing data (abdomen, brain, and leg) from six volunteers for training, while leaving out data from two volunteers for evaluation. Typical VFA T1-mapping acquisitions do not cover 30 flip angles, therefore each anatomical region from each volunteer in the training set was loaded 500 times, using different random subsets of flip angles as described in Table 2. Each training set (specific to each fold) was supplemented with 1,250,000 simulated signal curves, to ensure that all possible QMRI parameter combinations were seen during training. Both the in vivo and simulated signal curves were generated with varying sequence lengths and configurations of independent variables, following the procedure outlined in Section 2.3.1 and Table 2. In each batch, in vivo signal curves were sampled with a 6:1 probability compared to simulated signal curves. Training lasted for 750 epochs, with each epoch consisting of 100 data batches of size 64. The model trained on VFA T1-mapping simulation data was taken as initialization. The learning rate started at $2.5 \cdot 10^{-5}$, and halved at epochs 250, 350, 450, 550 and 650. Optimization was performed using the Adam optimizer (Kingma, 2015). Absolute and relative error tolerance of the adaptive ODE solver, used to compute the hidden state in Eq. (4) were set to 0.00001 and 0.001 respectively and halved at epochs 300, 400, 500 and 600.

Assessing the accuracy of QMRI parameter estimation methods in vivo is challenging due to the absence of ground truth for QMRI parameters. To establish reference values, an LSQ fit using all 30 flip angles was used as the reference standard. Subsequently, both LSQ fitting and NCDE-based parameter estimations were performed on input sequences containing smaller subsets of flip angles, as detailed in Table S3, which were generated according to the sampling protocol in Table 2. Within each volunteer, the sampling of different sequence lengths was initialized with the same random seed to ensure similarity across subsets of different length. Errors (as defined in Eq. (10)) and squared errors (as defined in Eq. (11)) between the reference standard and estimates were calculated per signal curve within a foreground mask, which excluded data with only zero values. The mean error and mean squared error were computed for each volunteer and anatomical region (abdomen, brain, and leg), and box plots were used to visualize the distribution of these metrics, grouped by anatomical region.

## 3. **Results**

### 3.1 Simulations

#### 3.1.1 VFA T1-mapping

Across all SNR levels < 200, NCDE-based parameter estimation consistently had a lower mean squared error, lower median (except for SNR = 150) and lower 75th percentile of the squared error than LSQ fitting at estimating the relaxation constant $T1$ (Fig. 2). However, at the highest SNR levels (SNR $\geq$ 200), LSQ fitting performed better than NCDE-based parameter estimation in estimating $T1$. Both NCDEs and LSQ fitting had more accurate and precise estimates when the input sequence became longer (Fig. 2). For all sequence lengths, NCDEs had a lower mean squared error than LSQ fitting in estimating the relaxation constant $T1$. At low SNR levels, NCDE-based parameter estimation yielded a decrease in interquartile range (IQR) of the $T1$ error compared to LSQ fitting without increasing the bias, as illustrated in Fig. S1.

NCDEs estimated QMRI parameters at a rate of 310 VFA T1-mapping curves per second, compared to 277 curves per second for LSQ fitting.

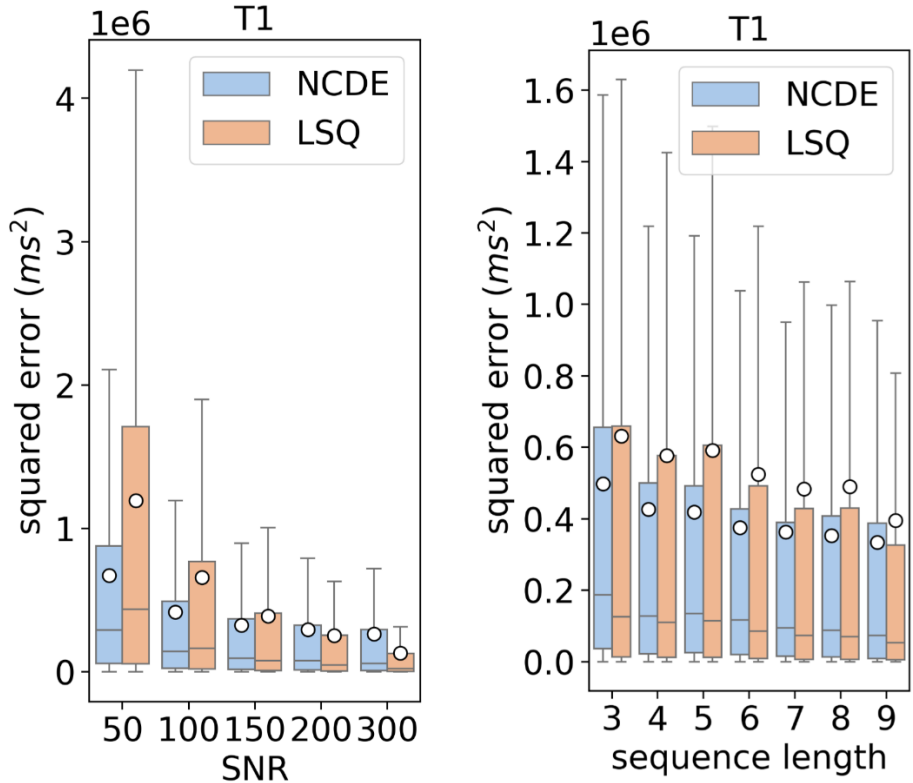

*Fig. 2: box plots of squared errors of estimated T1 parameters as a function of SNR (left) and as a function of sequence length (right) for NCDE (blue) and LSQ (orange). The white circle represents the mean squared error.*

### 3.1.2 IVIM MRI

At low SNR levels (SNR $\leq$ 20), NCDEs had lower mean squared error and median of the squared error than LSQ fitting across all IVIM MRI parameters (Fig. 3). At high SNR levels (SNR $\geq$ 40), LSQ improved estimates of the diffusion coefficient ($D$) and perfusion fraction ($f$) compared to NCDEs, although NCDEs retained better estimates of the pseudo-diffusion coefficient ($D^*$). Both NCDEs and LSQ fitting had more accurate and precise estimates when the input sequence became longer (Fig. 4). For almost all sequence lengths, NCDEs had lower mean squared error and lower 75th percentile of the squared error than LSQ fitting in estimating IVIM MRI parameters. At low SNR levels, NCDE-based parameter estimation yielded a decrease in IQR of the error across IVIM MRI parameters compared to LSQ fitting without increasing the bias (Fig. S2). NCDEs estimated QMRI parameters for 133 IVIM curves per second, compared to 140 curves per second for LSQ fitting.

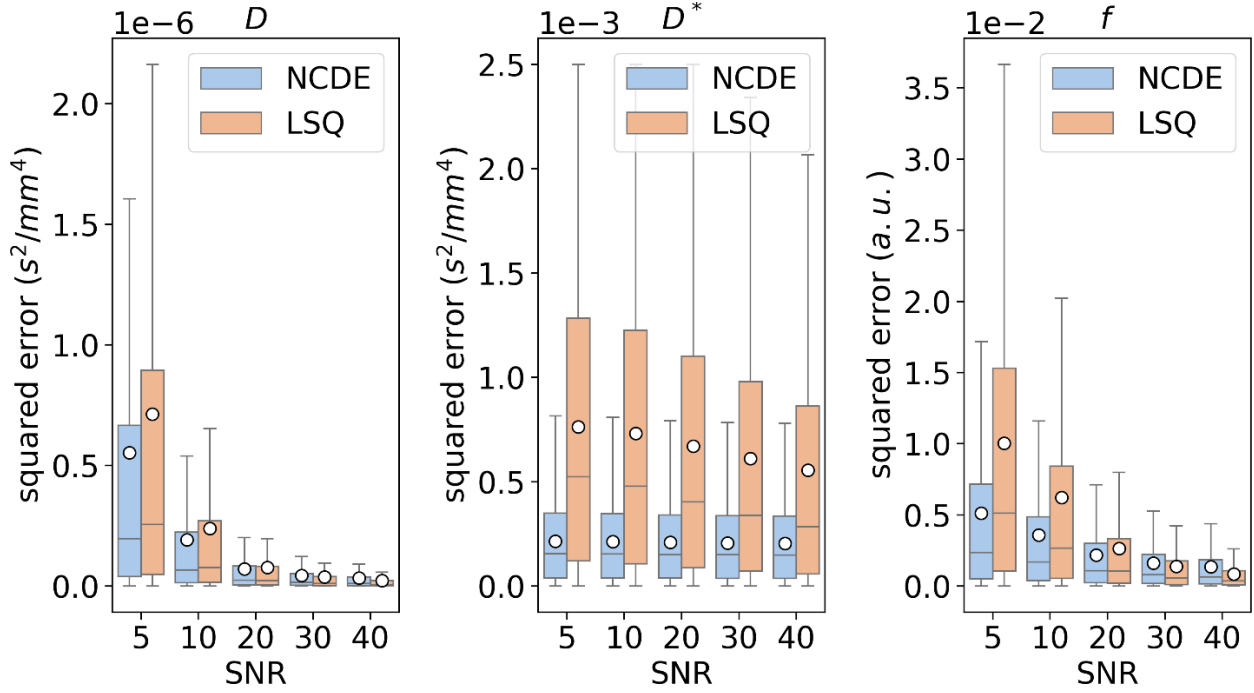

*Fig. 3: box plots of squared errors of estimated IVIM MRI parameters as a function of SNR for NCDE (blue) and LSQ (orange). The white circle represents the mean squared error.*

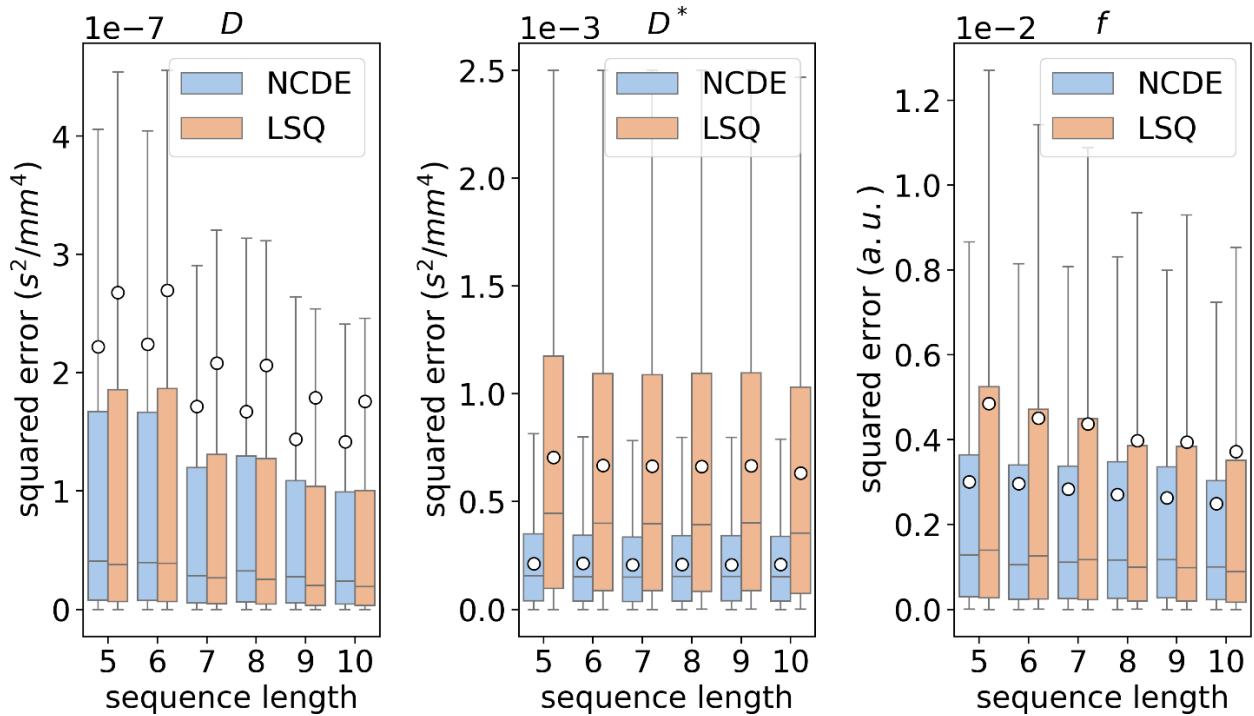

*Fig. 4: box plots of squared errors of estimated IVIM MRI parameters as a function of sequence length for NCDE (blue) and LSQ (orange). The white circle represents the mean squared error.*

### 3.1.3 DCE-MRI

At almost all SNR levels (SNR $\geq$ 10), NCDEs had a lower mean squared error than LSQ fitting across all DCE-MRI parameters (Fig. 5). In almost all of those cases, NCDEs also showed a lower 75$^{th}$ percentile of the squared error. At very low SNR (SNR = 5), LSQ had a lower mean squared error than NCDEs in estimating the reflux rate ($k_e$) and the fractional volume of the extravascular extracellular space ($v_e$), while NCDEs had a lower mean squared error in estimating the fractional volume of plasma ($v_p$). Both NCDEs and LSQ fitting had more accurate and precise estimates when the input sequence became longer (Fig. 6). For almost all sequence lengths, NCDEs had a lower mean squared error and lower 75th percentile of the squared error than LSQ fitting in estimating the DCE-MRI parameters. Across all SNR levels, NCDE-based parameter estimation yielded a comparable IQR and bias of the error across DCE-MRI parameters compared to LSQ fitting, as illustrated in Fig. S3. NCDEs estimated QMRI parameters for 60 DCE-MRI curves per second, compared to 27 curves per second for LSQ fitting.

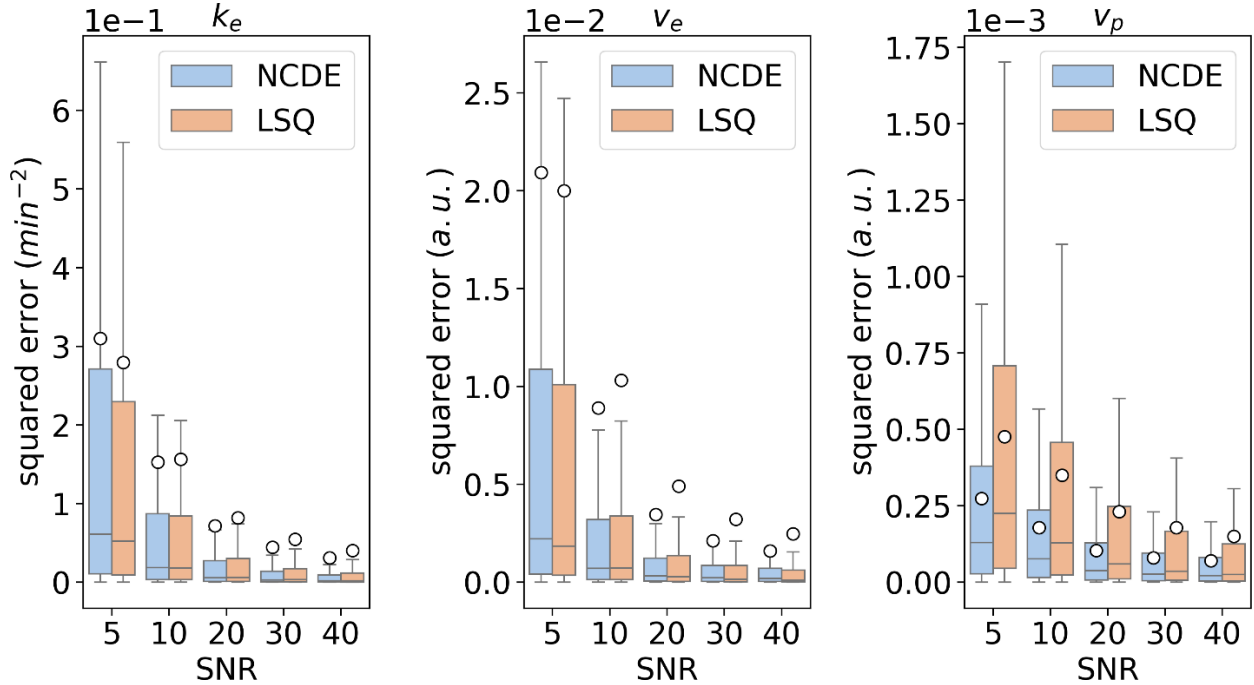

*Fig. 5: box plots of squared errors of estimated DCE-MRI parameters as a function of SNR for NCDE (blue) and LSQ (orange). The white circle represents the mean squared error.*

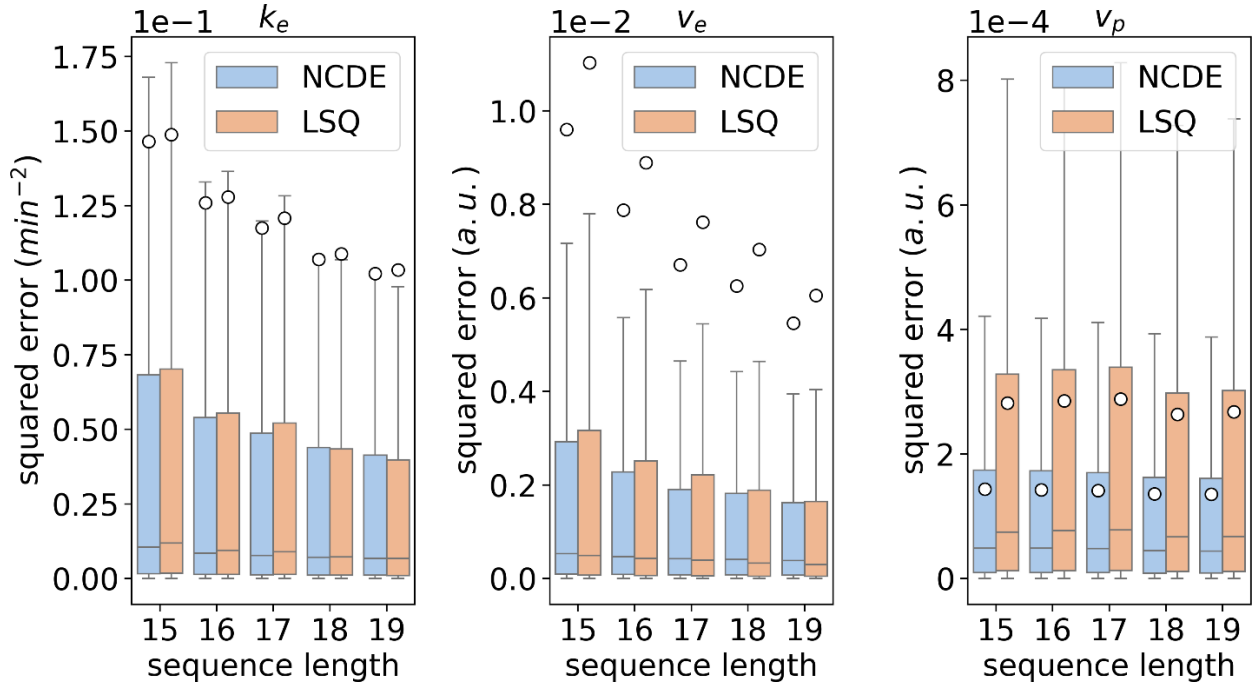

*Fig. 6: box plots of squared errors of estimated DCE-MRI parameters as a function of sequence length for NCDE (blue) and LSQ (orange). The white circle represents the mean squared error.*

### 3.2 In vivo data

NCDE-based parameter estimation resulted in more precise and more accurate QMRI parameter maps in vivo, especially in the abdomen and leg. Figs. 7-9 show typical examples of the QMRI parameter maps from NCDE-based parameter estimation and LSQ fitting, Figs. S4-S6 show their corresponding difference maps. In the abdomen, the anatomical region where T1 values are hardest to estimate, the NCDE better approximated the reference T1 values in noisy parts of the image than LSQ (Fig. 7, at arrows). For both NCDE-based parameter estimation and LSQ fitting with a subset of flip angles the delineation of different organs (kidney, liver, renal arteries) became less clear compared to the reference T1 map. In the anatomical region with the highest SNR, the brain (Fig. 8), the contrast between different tissues (grey matter, white matter, cerebrospinal fluid) was similar among NCDE-based parameter estimation and LSQ fitting with a subset of flip angles and the reference T1 map. For the legs, the contrast between different tissues (bone, muscle) was similar among NCDE-based parameter estimation and LSQ fitting with a subset of flip angles and the reference T1 map, but the NCDE-based QMRI parameter maps underestimated the T1 values in some areas of high T1 (Fig. 9, at arrows).

In abdominal and leg T1 maps, NCDE-based parameter estimation had a lower mean, median, 25th and 75th percentile of the mean squared error per volunteer over 3, 4 and 5 flip angles (Fig. 10) than LSQ fitting. In brain T1 maps, NCDE-based parameter estimation showed lower mean, median and 25th percentile of the mean squared error per volunteer compared to LSQ fitting for 3, 4 and 5 flip angles, while LSQ showed 75th percentile of the mean squared errors per volunteer at 5 flip angles. The NCDE had no problem dealing with varying amounts of input data as the squared errors decreased with increasing sequence length. NCDE-based parameter estimation yielded a smaller bias of the error compared to LSQ fitting across sequence length for data of the leg and abdomen, but a larger bias for brain data, as illustrated in Fig. S7.

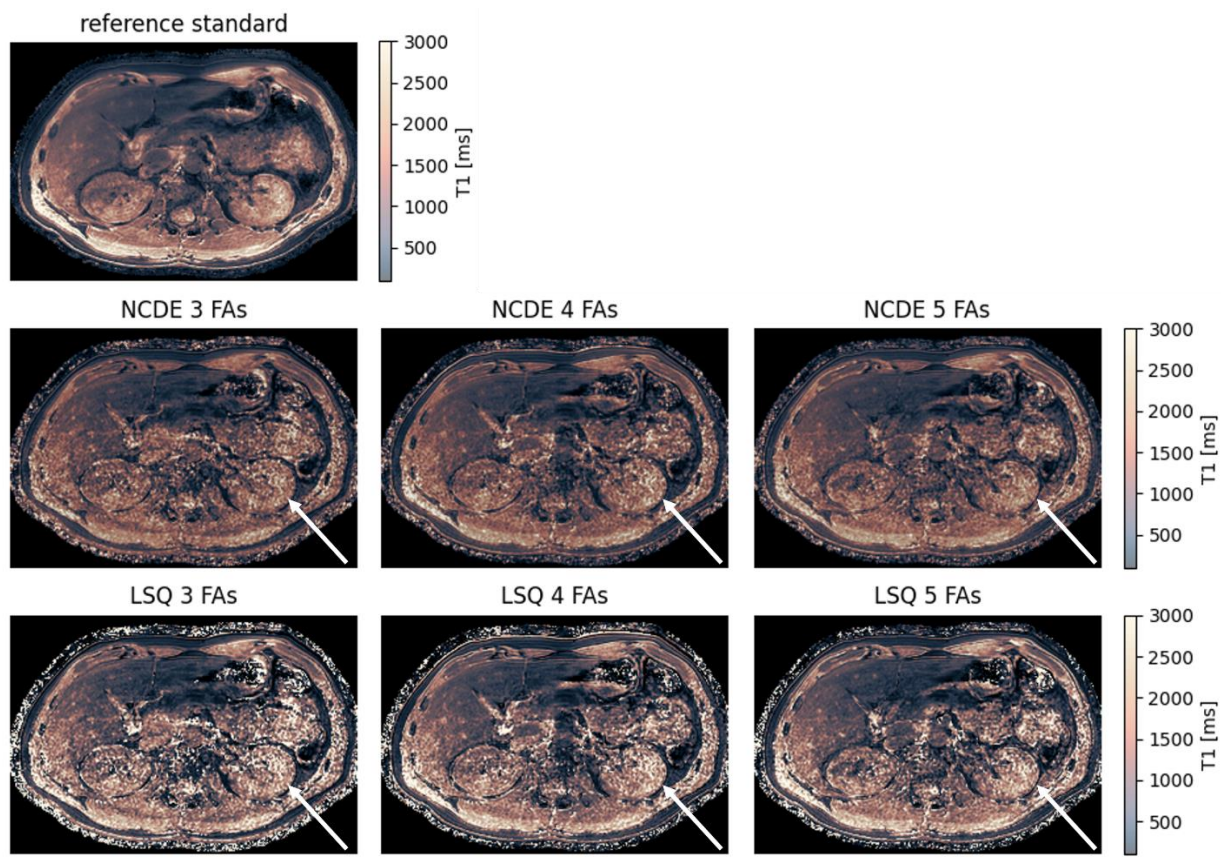

Fig. 7: comparison of NCDE-based and LSQ-based abdominal T1 parameter maps based on subsampled data: abdomen reference T1 parameter map (top row) and estimated T1 parameter maps for NCDE (middle row) and LSQ (bottom row). Arrows indicate a region (left kidney) where T1 parameter maps were degraded by noise, affecting LSQ fitting parameter maps more than NCDE-based parameter maps. Flip angles used for the evaluation at 3, 4 and 5 flip angles were [1, 6, 29]º, [1, 6, 15, 29]º and [1, 6, 10, 20, 29]º, respectively.

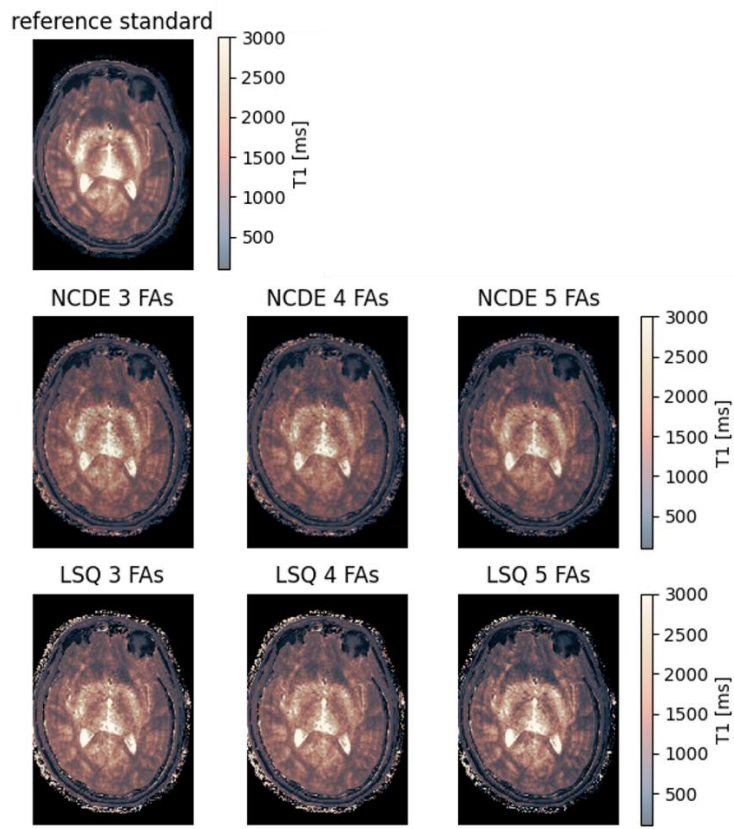

Fig. 8: comparison of NCDE-based and LSQ-based brain T1 parameter maps based on subsampled data: brain reference T1 parameter map (top row) and estimated T1 parameter maps for NCDE (middle row) and LSQ (bottom row). Flip angles used for the evaluation at 3, 4 and 5 flip angles were: [1, 9, 20]º, [1, 9, 10, 20]º and [1, 7, 9, 14, 20]º, respectively.

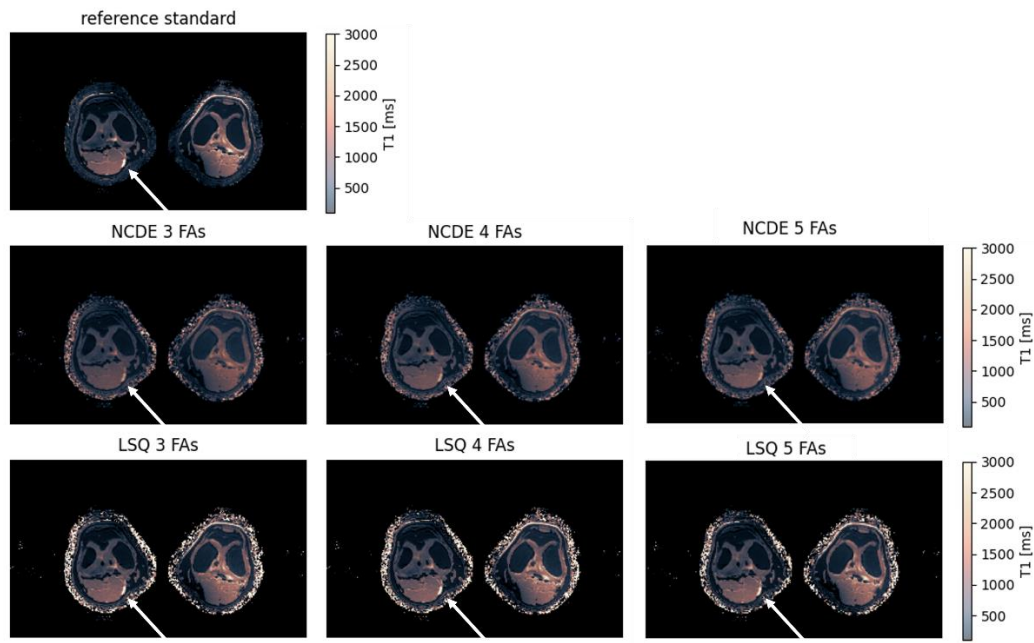

Fig. 9: *comparison of NCDE-based and LSQ-based T1 parameter maps of the legs based on subsampled data: leg reference T1 parameter map (top row) and estimated T1 parameter maps for NCDE (middle row) and LSQ (bottom row). Arrows indicate a region of high T1 values, which was better estimated by LSQ fitting than by NCDE-based parameter estimation. Flip angles used for the evaluation at 3, 4 and 5 flip angles were [1, 9, 25]º, [1, 9, 13, 25]º and [1, 9, 10, 17, 25]º, respectively.*

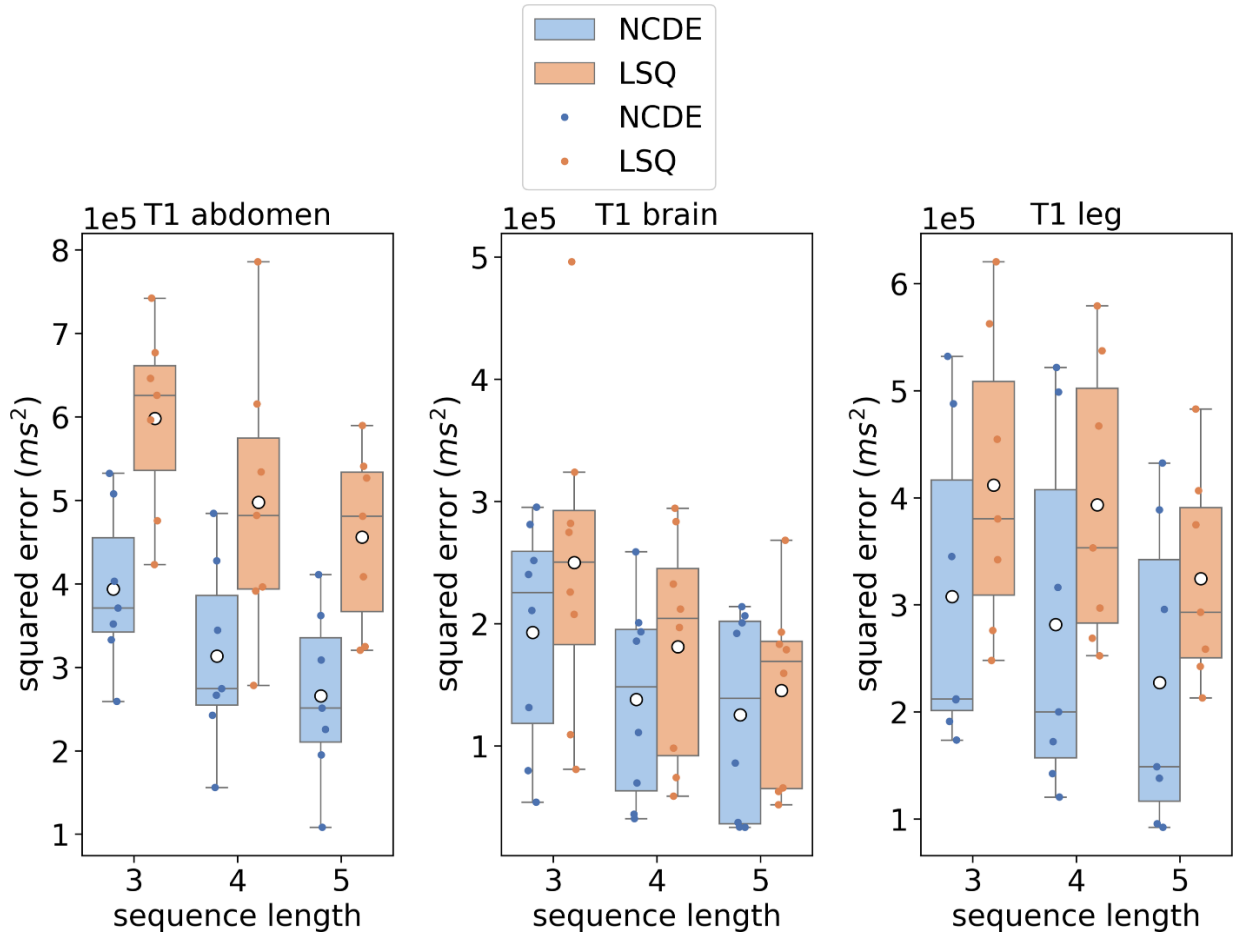

*Fig. 10: squared error of the estimated T1 parameter as a function of sequence length, per anatomical region. A box plot of the squared error per volunteer and per anatomical region between reference T1 parameter map and estimated T1 parameter maps for NCDE (blue) and LSQ (orange), the blue and orange dots represent the mean squared errors for each anatomical region per volunteer. Flip angles used for the evaluation at 3, 4 and 5 flip angles can be found in Table 3.*

## 4. Discussion

In this study, we show that NCDEs can function as a generic tool for the accurate prediction of QMRI parameters, regardless of the QMRI sequence length, the configuration of independent variables, or the QMRI forward model used. To the best of our knowledge, we are the first to propose a general solution for QMRI parameter estimation using deep learning with irregularly sampled data across different QMRI forward models, without retraining for different acquisition protocols. Our NCDE outperformed LSQ fitting for a broad range of settings, especially in the presence of uncertainty (low SNR and shorter input sequences). We are convinced that with NCDEs, we have solved one of the main challenges for using deep learning for QMRI parameter estimation in a broader clinical and research setting.

In simulations, NCDEs achieve lower mean squared error than LSQ fitting for all QMRI forward models when SNR levels are low. However, at high SNR, LSQ fitting remains the better method for estimating most QMRI parameters. NCDEs improve parameter estimation compared to LSQ fitting by reducing the IQR in estimation errors without an increase in bias. In vivo, NCDEs achieve lower mean squared errors than LSQ fitting in abdominal and leg VFA T1-mapping but show smaller improvements in the brain, where higher SNR and higher image quality make QMRI parameters easier to estimate. The difference in relative performance across anatomical regions suggests that NCDE-based QMRI parameter estimates are more accurate than LSQ fitting under conditions of high uncertainty, similar to the behavior that is observed in simulations. This work highlights the potential advantages of NCDE-based parameter estimation over LSQ fitting, particularly in challenging anatomical regions and high-uncertainty conditions, suggesting that NCDEs may offer a more robust approach for reliable QMRI.

With high SNR and densely sampled input sequences, the estimation of QMRI parameters is a trivial problem for which LSQ fitting performs sufficiently well. Unfortunately, such conditions are rare in clinical practice, which highlights that there is a need for suitable alternatives to LSQ fitting. Over the last 30 years, alternatives to LSQ fitting have been described in literature, including Bayesian inference (Barbieri et al., 2016; Baselice et al., 2016; Lofstedt et al., 2020; Neil, 1993; Scalco et al., 2022; Spinner et al., 2021; While, 2017) and neural networks (Bliesener et al., 2020; Epstein et al., 2024; Kaandorp et al., 2021; Ottens et al., 2022; Ulas et al., 2018; Vasylechko et al., 2022). The methods presented in these studies reduce the variation in estimation errors at low SNR levels, but are difficult to implement in a broader clinical context, for varying reasons. Bayesian inference is computationally intensive, time consuming and dependent on the choice of the prior distribution. Fully connected neural networks and convolutional neural networks are faster and learn prior distributions implicitly, but require a specifically trained neural network for each configuration of independent variables, creating considerable challenges for implementation and limiting reproducibility across different settings. Recurrent neural networks can handle variable length input sequences, but place assumptions on the regularity of the sampling interval which are often not met in quantitative MRI. Moreover, as training networks is a stochastic process, the need for retraining introduces an additional variation (Kaandorp et al., 2021) and hinders quality control. With the implementation of NCDEs, we aim to overcome most of these limitations. NCDEs are versatile networks that exhibit robustness to varied acquisition protocols, allowing end-users to directly apply a pre-trained, validated model, broadening access to QMRI parameter estimation for researchers and clinicians.

Although NCDEs have great potential, the NCDE-based parameter estimation method also has some specific requirements that may limit its applicability. For example, NCDE-based parameter estimation improved performance with input sequence normalization. In our implementation, this necessitated the acquisition of low flip angle images in VFA T1-mapping or images without diffusion weighting in IVIM MRI. For other implementations, other normalization techniques may be used. Further, we encountered difficulties when training the current NCDE implementation on longer sequences (>30 datapoints), leading to long training times and not finding adequate parameterizations of the NCDE in the training process. This is a known aspect of NCDEs (Kidger, 2022), with proposed solutions (Morrill, 2021; Walker, 2024) whose application in QMRI parameter estimation (e.g. densely sampled DCE-MRI) fall outside the scope of this work. Finally, we demonstrated that NCDEs were faster in estimating parameters in VFA T1-mapping and DCE-MRI, but LSQ fitting only required CPUs while NCDEs required GPUs.

One limitation of our in vivo results is that the comparison between NCDEs and LSQ fitting is constrained by the absence of definitive ground truth values. While we attempted to establish a robust reference standard using 30 flip angles, the resulting values remained susceptible to noise, image corruption, and fitting constraints.

The field of deep learning-based QMRI parameter estimation is still in its early stages, leaving many questions unanswered. Although several studies have shown improvements of deep learning-based QMRI parameter estimation over the current state-of-the-art (Barbieri et al., 2020; Kaandorp et al., 2021; Ottens et al., 2022; Vasylechko et al., 2022), the underlying mechanisms for these improvements compared to LSQ fitting remain obscure. The limitations of the LSQ fitting algorithm are clear: it lacks prior information on expected distributions, does not enforce spatial coherence, is highly sensitive to outliers and noise and the complexity of QMRI forward models reduces the likelihood of reaching the global optimum in the optimization process. However, these very shortcomings contribute to its strength as LSQ fitting has become widely successful due to its simplicity, usability and predictability. NCDEs use a different approach, directly inferring the QMRI parameters from the signal curve without relying on an iterative process. This approach reduces their vulnerability to noise and the challenges posed by complex optimization landscapes. NCDEs have the potential to rival LSQ fitting in terms of usability and predictability, while offering advantages in finding QMRI parameter optima, ensuring spatial coherence and improving noise robustness and speed. Finally, the flexibility of NCDEs not only extends to a variety of MRI techniques, including pharmacokinetic DCE-MRI models, diffusion tensor imaging, and MR fingerprinting, but also to other modalities like positron emission tomography and ultrasound elastography, positioning NCDEs as versatile tools across imaging modalities.

In our eyes, the main advantage of NCDEs is that it allows training one network per QMRI technique that can be used around the world for any dataset, regardless of acquisition protocol or anatomical region. This would have several advantages, including easy implementation, quality assurance of the model and comparable results across various sites. Our short-term goal is to train such models for IVIM MRI, DCE-MRI and VFA T1-mapping and share them with the community. Ultimately, to train such models with direct applicability in research and clinical settings, a large and varied dataset of quantitative data from multiple sites is required, with data from multiple vendors, field strengths, anatomical regions and acquisition protocols. To achieve this, we make an appeal to our readers. If the reader is willing to contribute to this dataset, we would appreciate if the reader could reach out to the authors.

## 5. Conclusion

NCDEs can function as a generic tool for accurate prediction of quantitative MRI parameters, irrespective of sequence length or quantitative MRI forward model. Overcoming these two hurdles opens up the use of neural networks for quantitative MRI parameter estimation to a broad audience of researchers and clinicians.

## 6. Acknowledgements

This work was funded by the KWF Dutch Cancer Society (KWF-UVA 2021.13785, OG-C and DK) and the Swiss National Science Foundation (32003B_176229/1, SB).

## 8. **Supplementary materials**

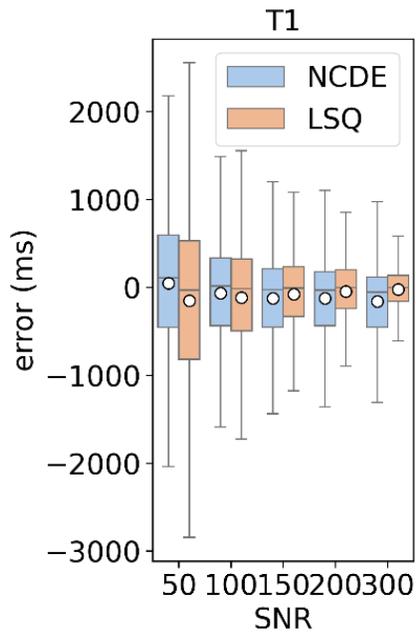

*Fig. S1: box plot of error of estimated T1 parameters (y-axis) as a function of SNR (x-axis) for NCDE (blue) and LSQ (orange). The white circle represents the mean error.*

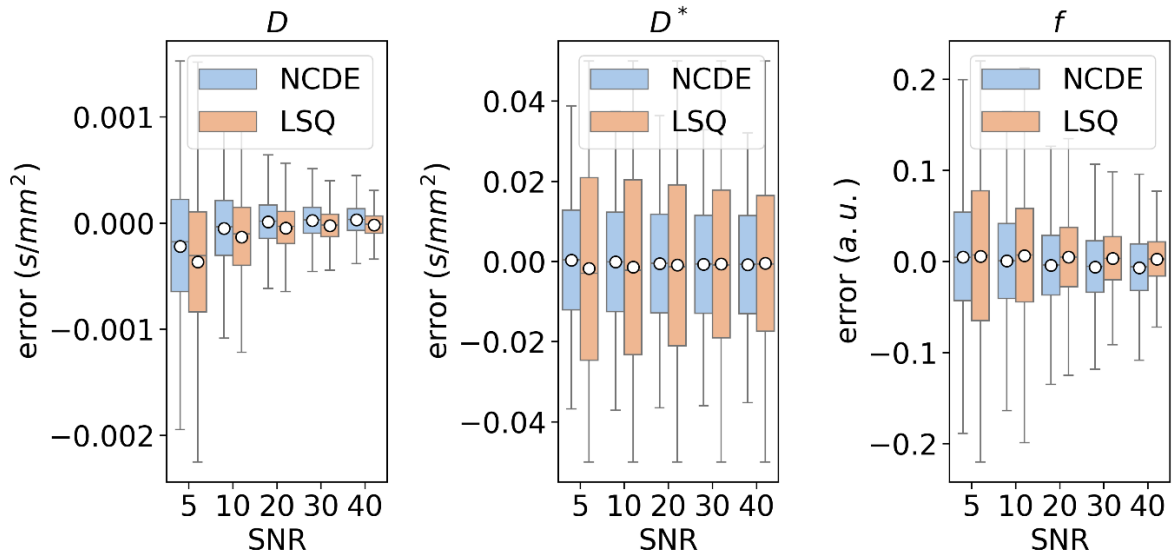

*Fig. S2: box plot of error of estimated IVIM MRI parameters (y-axes) as a function of SNR (x-axes) for NCDE (blue) and LSQ (orange). The white circle represents the mean error.*

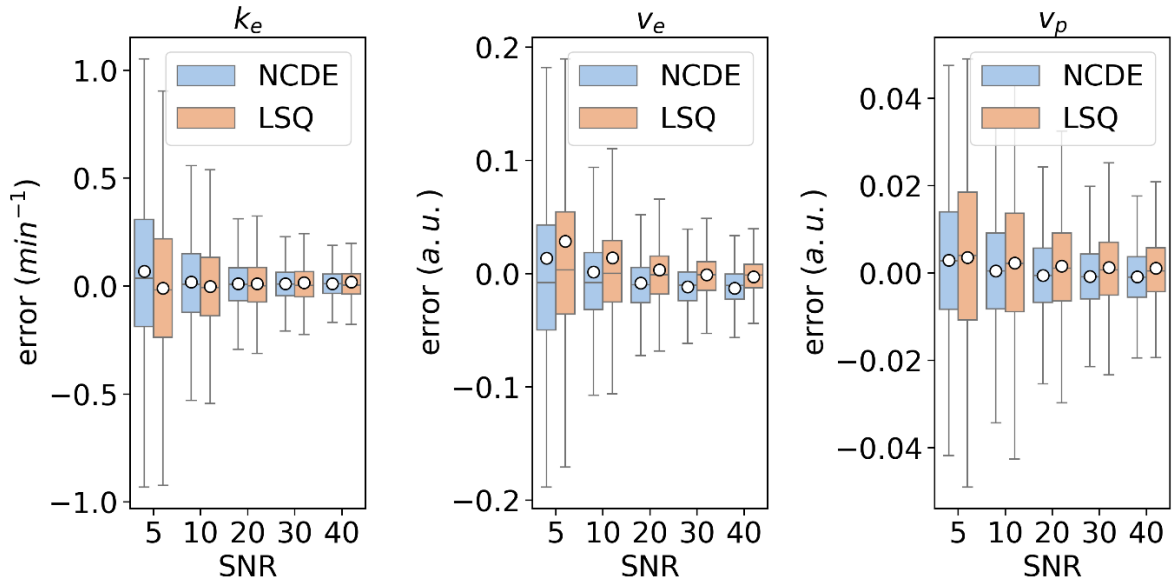

Fig. S3: box plot of error of estimated DCE-MRI parameters (y-axes) as a function of SNR (x-axes) for NCDE (blue) and LSQ (orange). The white circle represents the mean error.

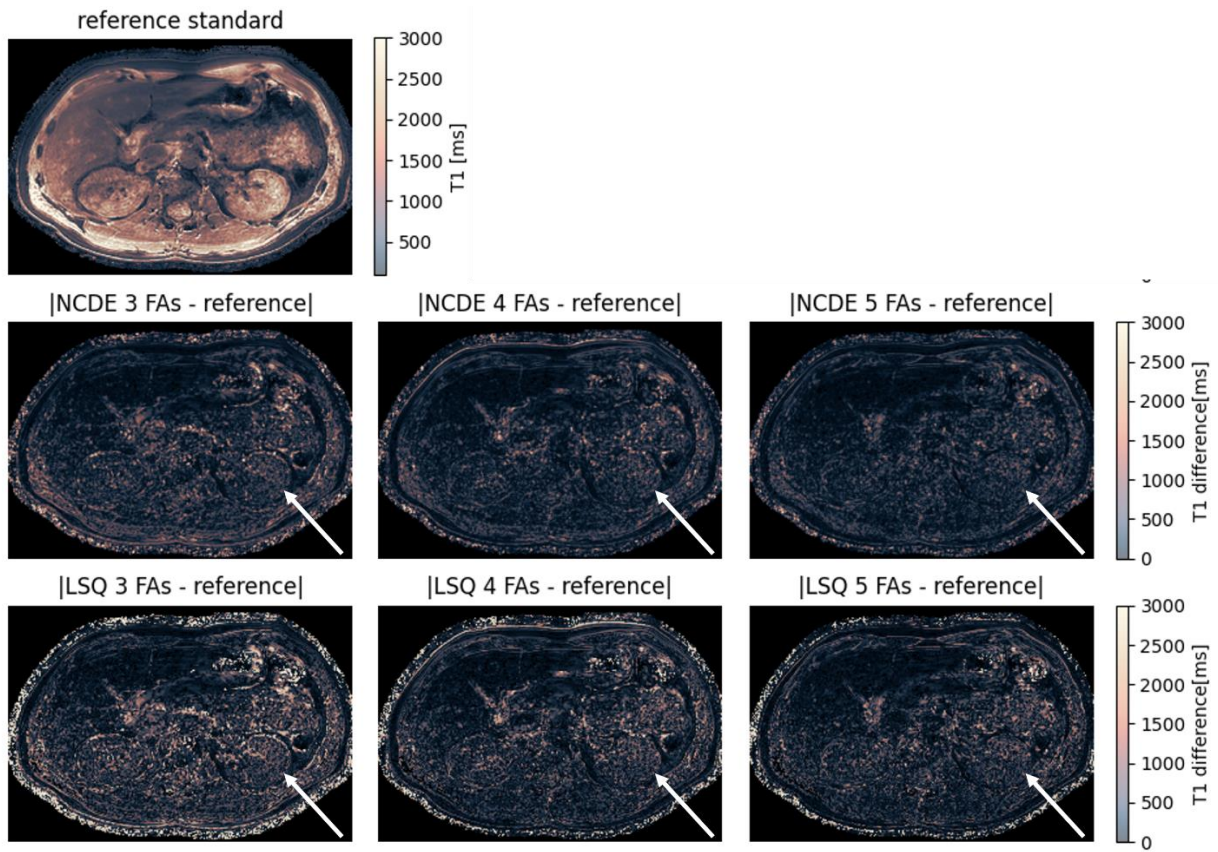

Fig. S4: comparison of difference maps between reference standard and NCDE-based and LSQ-based abdominal T1 parameter maps based on subsampled data: abdomen reference T1 parameter map (top row) and T1 difference maps for NCDE (middle row) and LSQ (bottom row). Arrows indicate a region (left kidney) where parameter maps were degraded by noise, affecting LSQ fitting

*parameter maps more than NCDE-based parameter maps. Flip angles used for the evaluation at 3, 4 and 5 flip angles were [1, 6, 29]º, [1, 6, 15, 29]º and [1, 6, 10, 20, 29]º, respectively.*

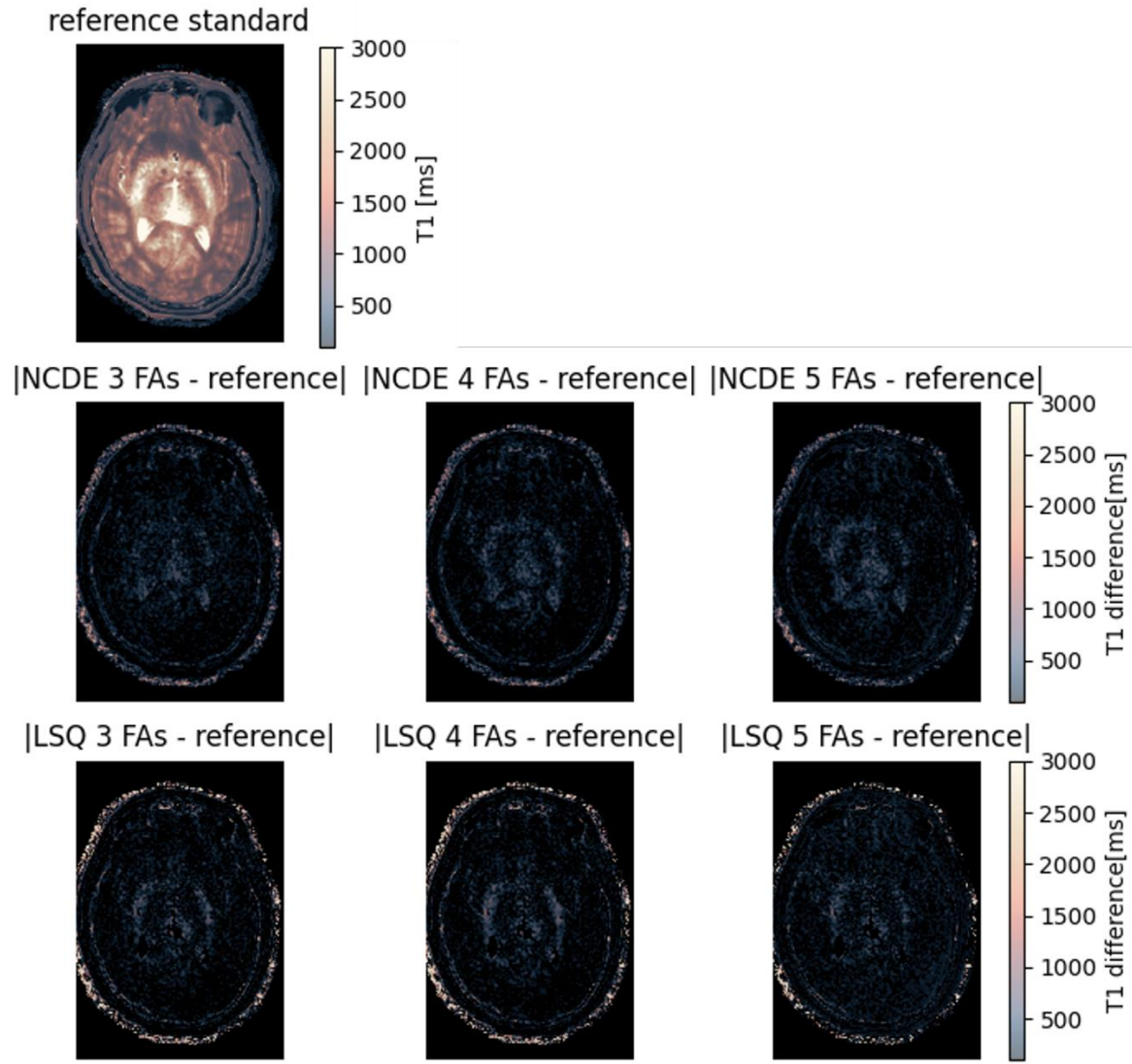

*Fig. S5: comparison of difference maps between reference standard and NCDE-based and LSQ-based brain T1 parameter maps based on subsampled data: brain reference T1 parameter map (top row) and estimated T1 difference maps for NCDE (middle row) and LSQ (bottom row). Flip angles used for the evaluation at 3, 4 and 5 flip angles were [1, 9, 20]º, [1, 9, 10, 20]º and [1, 7, 9, 14, 20]º, respectively.*

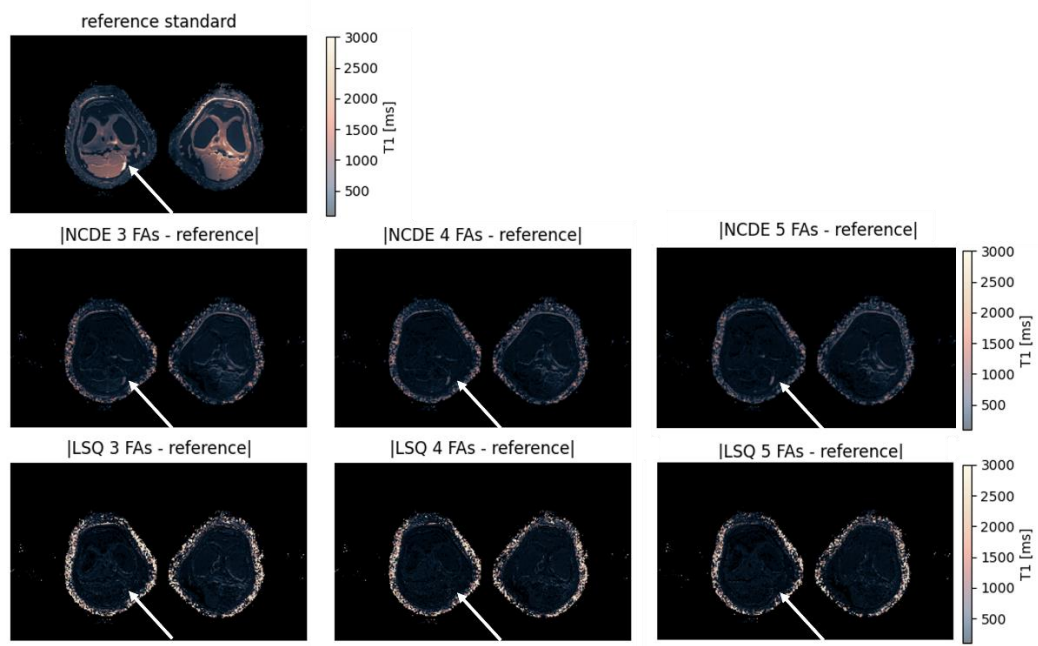

*Fig. S6: comparison of difference maps between reference standard and NCDE-based and LSQ-based brain T1 parameter maps based on subsampled data: leg reference T1 parameter map (top row) and estimated T1 difference maps for NCDE (middle row) and LSQ (bottom row). Arrows indicate a region of high T1 values, which was better estimated by LSQ fitting than by NCDE-based parameter estimation. Flip angles used for the evaluation at 3, 4 and 5 flip angles were [1, 9, 25]º, [1, 9, 13, 25]º and [1, 9, 10, 17, 25]º, respectively.*

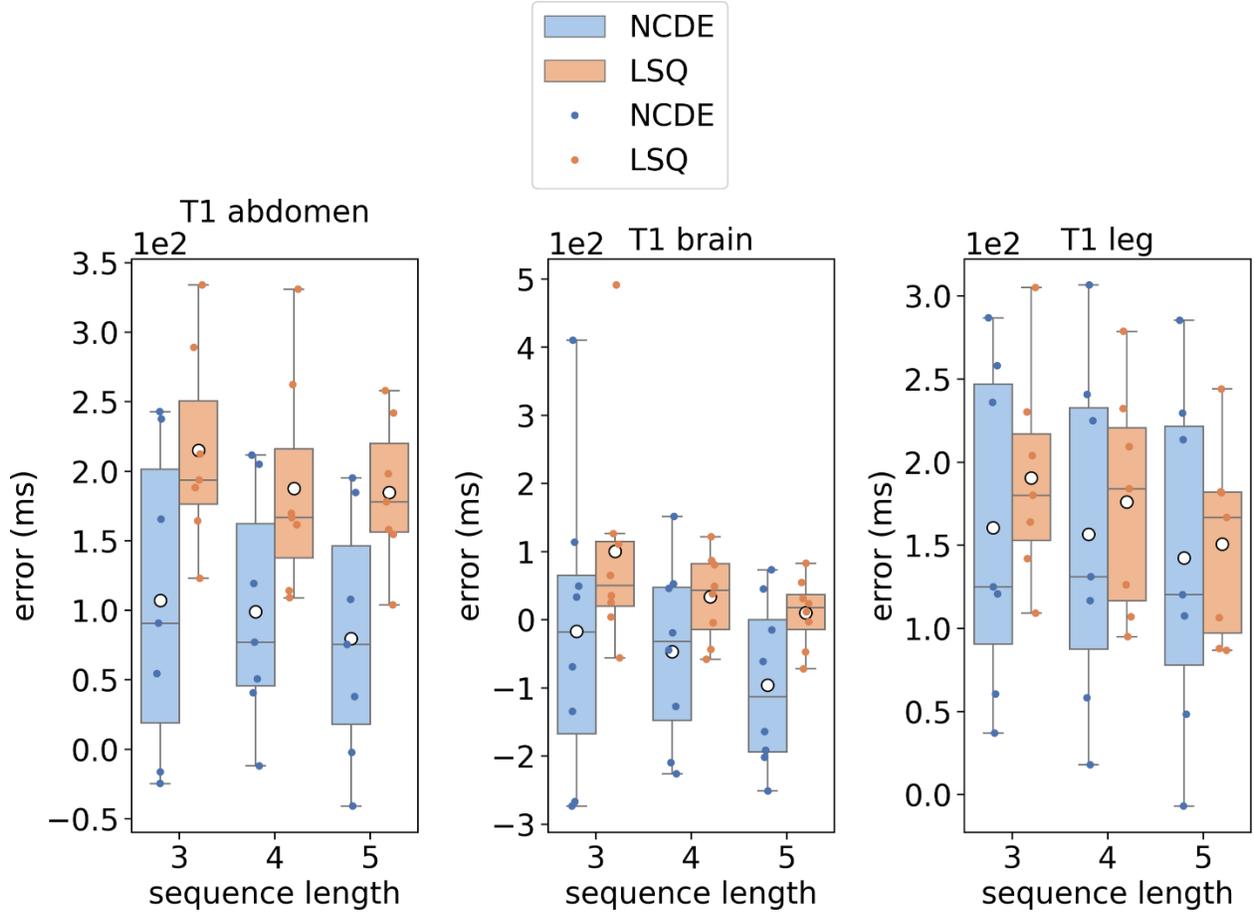

Fig. S7: error of the estimated T1 parameter as a function of sequence length, per anatomical region. A box plot of the error per volunteer and per anatomical region between reference T1 parameter map and estimated T1 parameter maps for NCDE (blue) and LSQ (orange). Flip angles used for the evaluation at 3, 4 and 5 flip angles can be found in Table 3.

| Layer Type | Output Shape | Number of model parameters |
|---|---|---|
| $l_\phi^1$ | | |
| └ Linear | [batch size, 1000] | 4,000 |
| $f_\phi$ | | |
| └ Linear | [batch size, 128] | 128,128 (recursive) |
| └ Rectified Linear Unit | [batch size, 128] | -- |
| └ Linear | [batch size, 256] | 33,024 (recursive) |
| └ Rectified Linear Unit | [batch size, 128] | -- |
| └ Linear | [batch size, 512] | 131,584 (recursive) |
| └ Rectified Linear Unit | [batch size, 128] | -- |
| └ Linear | [batch size, 256] | 131,328 (recursive) |
| └ Rectified Linear Unit | [batch size, 128] | -- |
| └ Linear | [batch size, 128] | 32,896 (recursive) |
| └ Rectified Linear Unit | [batch size, 128] | -- |
| └ Linear | [batch size, 3000] | 387,000 (recursive) |
| └ Tanh | [batch size, 3000] | -- |
| $l_\phi^2$ (parallel for each estimated QMRI parameter) | | |
| └ Linear | [batch size, 1000] | 1,001,000 (per estimated QMRI parameter) |
| └ 1D batch normalization | [batch size, 1000] | 2,000 (per estimated QMRI parameter) |
| └ Rectified Linear Unit | [batch size, 1000] | -- |
| └ Linear | [batch size, 1000] | 1,001,000 (per estimated QMRI parameter) |
| └ 1D batch normalization | [batch size, 1000] | 2,000 (per estimated QMRI parameter) |
| └ Rectified Linear Unit | [batch size, 1000] | - |
| └ Linear | [batch size, 1000] | 1,001,000 (per estimated QMRI parameter) |
| └ 1D batch normalization | [batch size, 1000] | 2,000 (per estimated QMRI parameter) |
| └ Rectified Linear Unit | [batch size, 1000] | -- |
| └ Linear | [batch size, 1000] | 1,001,000 (per estimated QMRI parameter) |
| └ 1D batch normalization | [batch size, 1000] | 2,000 (per estimated QMRI parameter) |
| └ Rectified Linear Unit | [batch size, 1000] | -- |
| └ Linear | [batch size, 1000] | 1,001,000 (per estimated QMRI parameter) |
| └ 1D batch normalization | [batch size, 1000] | 2,000 (per estimated QMRI parameter) |
| └ Rectified Linear Unit | [batch size, 1000] | -- |
| └ Linear | [batch size, 1] | 1,001 (per estimated QMRI parameter) |
| Total model parameters: | | for n estimated QMRI parameters: 847960 + 5015000*n |
| Trainable model parameters: | | for n estimated QMRI parameters: 847960 + 5015000*n |
| Non-trainable model parameters: | | 0 |

*Table S1: details of the architecture of the NCDE. The readout is designed as a separate parallel track for each QMRI parameter. A description of $l_0^1$, $f_0$ and $l_0^2$ can be found in section 2.2.*

| Anatomical region | Brain | Abdomen | Leg |
|---|---|---|---|
| Sequence type | Spoiled gradient echo | Spoiled gradient echo | Spoiled gradient echo |
| Repetition time (ms) | 4.61 - 5.1 | 5.87 – 6 | 5.87 - 15 |
| Echo time (ms) | 1.6 -1.6 | 1.5 | 1.6 |
| Flip angle (degree) | 1-30 | 1-30 | 1-30 |
| Range of reconstructed pixel spacing (frequency x phase x slices) (mm) | (0.86, 0.86, 1.5) - (0.90, 0.90, 2) | (0.67, 0.67, 1.5) - (0.94, 0.94, 1.5) | (0.66, 0.66, 1.5) - (0.85, 0.85, 1.5) |
| Acquisition time (s) | 168 - 891 | 271 - 478 | 467 - 631 |
| Motion management | - | Breath hold per flip angle | - |

*Table S2: Acquisition settings of the MRI data acquired on a Philips Ingenia 3T system (Philips, Best, The Netherlands) for brain, abdomen and leg.*

| Volunteer number | Number of flip angles used in evaluation | Subset of flip angles used in evaluation | | |
|---|---|---|---|---|
| | | Brain | Abdomen | Leg |
| V1 | 3 | [1, 8, 30]º | [1, 10, 30]º | [1, 6, 25]º |
| | 4 | [1, 8, 16, 30]º | [1, 10, 16, 30]º | [1, 6, 13, 25]º |
| | 5 | [1, 8, 11, 20, 30]º | [1, 10, 11, 20, 30]º | [1, 6, 9, 17, 25]º |
| V2 | 3 | [1, 10, 24]º | [1, 10, 17]º | [1, 5, 22]º |
| | 4 | [1, 10, 13, 24]º | [1, 9, 10, 17]º | [1, 5, 12, 22]º |
| | 5 | [1, 9, 10, 16, 24]º | [1, 6, 10, 12, 17]º | [1, 5, 8, 15, 22]º |
| V3 | 3 | [1, 9, 17]º | [1, 4, 28]º | [1, 9, 25]º |
| | 4 | [1, 9, 10, 17]º | [1, 4, 15, 28]º | [1, 9, 13, 25]º |
| | 5 | [1, 6, 9, 12, 17]º | [1, 4, 10, 19, 28]º | [1, 9, 10, 17, 25]º |

| V4 | 3 | [1, 4, 19]º | N.A. | [1, 3, 17]º |
|----|---|-------------|------|-------------|
|    | 4 | [1, 4, 10, 19]º | N.A. | [1, 3, 9, 17]º |
|    | 5 | [1, 4, 7, 13, 19]º | N.A. | [1, 3, 6, 12, 17]º |
| V5 | 3 | [1, 9, 20]º | [1, 6, 29]º | [1, 5, 25]º |
|    | 4 | [1, 9, 10, 20]º | [1, 6, 15, 29]º | [1, 5, 13, 25]º |
|    | 5 | [1, 7, 9, 14, 20]º | [1, 6, 10, 20, 29]º | [1, 5, 9, 17, 25]º |
| V6 | 3 | [1, 3, 20]º | [1, 6, 17]º | N.A. |
|    | 4 | [1, 3, 10, 20]º | [1, 6, 9, 17]º | N.A. |
|    | 5 | [1, 3, 7, 14, 20]º | [1, 6, 7, 12, 17]º | N.A. |
| V7 | 3 | [1, 6, 24]º | [1, 3, 18]º | [1, 4, 20]º |
|    | 4 | [1, 6, 13, 24]º | [1, 3, 10, 18]º | [1, 4, 10, 20]º |
|    | 5 | [1, 6, 9, 16, 24]º | [1, 3, 7, 12, 18]º | [1, 4, 7, 14, 20]º |
| V8 | 3 | [1, 9, 30]º | [1, 9, 30]º | [1, 9, 22]º |
|    | 4 | [1, 9, 16, 30]º | [1, 9, 16, 30]º | [1, 9, 12, 22]º |
|    | 5 | [1, 9, 11, 20, 30]º | [1, 9, 11, 20, 30]º | [1, 8, 9, 15, 22]º |

*Table S3: description of in vivo data per organ and per volunteer of the subset of flip angles used for evaluation of the NCDE model and least squares fitting. Data originated from one study performed in our center, all acquired on a Philips Ingenia 3.0T.*

| Parameter | Value |
|-----------|-------|
| aB | 7.9785 |
| μB | 32.8855 |
| μM | 9.1868 |
| aR | 0.0482 |
| μR | 15.8167 |
| tR | 0.2533 |
| aE | 0.5216 |
| μE | 0.1811 |
| t0 | 0 |

*Table S4: description of the parameterized version of Cp as used in (Rata et al., 2016).*